\documentclass[12pt,a4paper,dvips,epsfig,openany]{article}
\usepackage{a4p,rotating}
\usepackage{epsfig,subfigure,amsmath}
\usepackage{cite,mcite}
\usepackage{graphicx}
\usepackage{fleqn}
\usepackage{hub_title,ifthen}
\usepackage{feynmf}

\journalname{Phys. Lett. B}
\preprint{99/06}
\graphicspath{{/l3/paper/example/figs/}}
\newcommand{\Zo}{$\mathrm{Z}^0$}

\newcommand{\Wpm}{$\mathrm{W}^\pm$}

\newlength{\capindent}
\newlength{\capwidth}
\newlength{\figwidth}
\newcommand{\icaption}[2][!*!,!]{\hspace*{\capindent}%
  \begin{minipage}{\capwidth}
    \ifthenelse{\equal{#1}{!*!,!}}%
      {\caption{#2}}%
      {\caption[#1]{#2}}
  \end{minipage}}
\begin{document}
\begin{titlepage}
\title{{\LARGE Measuring WWZ and WW$\gamma$ coupling constants with
    \Zo-pole data }}

\author{ Peter Moln\'ar and Martin Gr\"unewald }

%
%
\begin{abstract}
  
  Triple gauge boson couplings between \Zo, $\gamma$ and the W boson are
  determined by exploiting their impact on radiative corrections to
  fermion-pair production in e$^+$e$^-$ interactions at centre-of-mass
  energies near the \Zo-pole.  Recent values of observables in the
  electroweak part of the Standard model are used to determine the
  four parameters $\epsilon_1$, $\epsilon_2$, $\epsilon_3$ and
  $\epsilon_b$.  In a second step the results on the four $\epsilon$
  parameters are used to determine the couplings $\Delta g^1_Z$ and
  $\Delta\kappa_\gamma$.  For a wide range of scales, these indirect
  coupling measurements are more precise than recent direct
  measurements at LEP 2 and at the TEVATRON.  The Standard model
  predictions agree well with these measurements.
 
\end{abstract}

\end{titlepage}
%
%
\section{Introduction}

One of the most prominent goals of the LEP 2 program performed at the
Large Electron Positron Collider (LEP) is the precise measurement of
the couplings between the neutral electroweak bosons \Zo, $\gamma$
and the charged boson \Wpm~\cite{LEP2YRAC}. Analogous measurements were performed
at the TEVATRON measuring mainly the coupling between the photon and
the \Wpm.  These two measurements were the first ones which were
able to prove the non-Abelian character of the electroweak part of the
Standard model~\cite{standard_model}. Even more precise determinations
will be possible at
future hadron or electron-positron-collider.

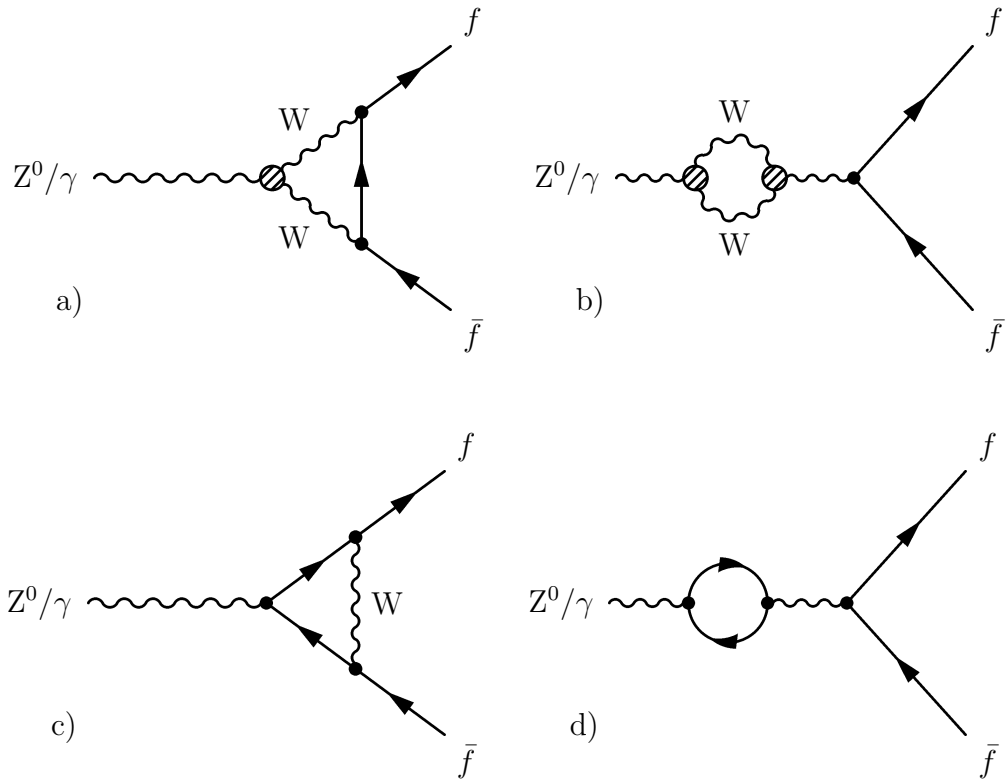
\begin{figure}[p]
\begin{center}
\begin{fmffile}{rb1}
a)
\begin{fmfgraph*}(150,100)
\fmfleft{i1}\fmfrightn{o}{2}
\fmflabel{\Zo/$\gamma$}{i1}
\fmflabel{$f$}{o2}
\fmflabel{$\bar{f}$}{o1}
\fmf{boson}{i1,v1}
\fmf{fermion}{o1,v2}
\fmf{fermion}{v3,o2}
\fmf{boson,label=W,l.side=left}{v2,v1,v3}
\fmffreeze
\fmf{fermion}{v2,v3}
\fmfdot{v2}\fmfdot{v3}\fmfblob{.06w}{v1}
\end{fmfgraph*}\hspace{1cm}
b)
\begin{fmfgraph*}(150,100)
\fmfleft{i1}\fmfrightn{o}{2}
\fmflabel{\Zo/$\gamma$}{i1}
\fmflabel{$f$}{o2}
\fmflabel{$\bar{f}$}{o1}
\fmf{phantom}{i1,v3}
\fmf{fermion}{o1,v3,o2}
\fmffreeze
\fmf{boson}{i1,v1}
\fmf{boson,left,tension=.5,label=W}{v1,v2,v1}
\fmf{boson}{v2,v3}
\fmfdot{v3}\fmfblob{.06w}{v1}\fmfblob{.06w}{v2}
\end{fmfgraph*} \\[2cm]
c)
\begin{fmfgraph*}(150,100)
\fmfleftn{i}{1}\fmfrightn{o}{2}
\fmflabel{\Zo/$\gamma$}{i1}
\fmflabel{$f$}{o2}
\fmflabel{$\bar{f}$}{o1}
\fmf{boson}{i1,v1}
\fmf{fermion}{o1,v2,v1,v3,o2}
\fmffreeze
\fmf{boson,label=W}{v2,v3}
\fmfdot{v1}\fmfdot{v2}\fmfdot{v3}
\end{fmfgraph*}\hspace{1cm}
d)
\begin{fmfgraph*}(150,100)
\fmfleft{i1}\fmfrightn{o}{2}
\fmflabel{\Zo/$\gamma$}{i1}
\fmflabel{$f$}{o2}
\fmflabel{$\bar{f}$}{o1}
\fmf{phantom}{i1,v3}
\fmf{fermion}{o1,v3,o2}
\fmffreeze
\fmf{boson}{i1,v1}
\fmf{fermion,left,tension=.5}{v1,v2,v1}
\fmf{boson}{v2,v3}
\fmfdot{v3}\fmfdot{v1}\fmfdot{v2}
\end{fmfgraph*}
\end{fmffile}

\end{center}
\caption{Radiative correction to  the decay width of the \Zo into
  fermions, \Zo$\rightarrow f\bar{f}$. This process is used to constrain
  the Higgs-boson and top-quark mass. Graphs a) and b) depend on the
  WWV coupling constants, while graphs c) and d) depend only on
  fermion to boson couplings.}
\label{fig:Z_ff}
\end{figure}

However, before the LEP 2 program with centre-of-mass energies above
the W-pair production threshold of about 161~GeV, LEP was running at
energies around the \Zo-pole at 91~GeV allowing to perform very
precise measurements of fermion pair production properties.  The
experiments at LEP-1 and also at SLAC measure radiative corrections to
the \Zo ff vertex.  These radiative corrections involve contributions with WWV (V=\Zo,
$\gamma$) vertices as shown in figure~\ref{fig:Z_ff} a) and b) and
WWV-independent contributions (figure~\ref{fig:Z_ff} c,d).
Therefore precise measurements of fermion-pair production allow the
determination of the WWV coupling constants.  This was noted already
in the beginning of the LEP era
\cite{DeRujula:1992se,hisz}.

The phenomenological effective Lagrangian of the WWZ and WW$\gamma$
vertices, respecting only Lorentz-invariance, contains 14 triple gauge
coupling constants (TGCs) as free parameters.  All of these can be
accommodated in the Standard Model requesting SU(2)$\times$U(1) gauge
invariance, if one considers higher dimensional SU(2)$\times$U(1)
gauge invariant operators. The neglect of higher dimensional operators
leads automatically to relations between TGCs. The model which is
discussed in the following neglects operators having a higher
dimension than six.  Loop corrections in this model lead to a
logarithmic divergence of low energy observables
\cite{DeRujula:1992se}.  However it was shown that three dimension-six
operators, that induce non-standard TGCs do not have this
property\cite{hisz}.  Assuming the existence of a light Higgs boson,
created by the Higgs-doublet field $\Phi$, one can apply a
linear realization of the SU(2)$\times$U(1) symmetry.  Then one obtains
in addition to the SM Lagrangian the following three terms~\cite{hisz}~:
\begin{align}
\Delta{\cal L}&=
\imath g'\frac{\Delta\kappa_\gamma-\cos^2\theta_W\Delta g^1_Z}{m^2_W}(D_\mu\Phi)^\dagger B^{\mu\nu}(D_\nu\Phi)+
\imath
g\frac{\cos^2\theta_W\Delta
  g^1_Z}{m^2_W}(D_\mu\Phi)^\dagger\vec{\tau}\cdot\vec{\hat{W}}^{\mu\nu}(D_\nu\Phi) \nonumber \\
&+\imath g\frac{\lambda_\gamma}{6m^2_W}\vec{\hat{W}}B^\mu_\nu\cdot(\vec{\hat{W}}B^\nu_\rho\times\vec{\hat{W}}B^\rho_\mu).
\end{align}

In this  model the TGC-relations are~:
\begin{align}
\Delta\kappa_\gamma&=\frac{\cos^2\theta_W}{\sin^2\theta_W}(\Delta\kappa_Z-\Delta
g^1_Z) ,\\
\lambda_\gamma&=\lambda_Z.
\end{align}
The remaining nine coupling constants are zero. The
SM predicts that all 14 parameters are zero. The TGCs $\Delta\kappa_V$
and $\Delta g^1_V$ parametrise the difference of  $g^1_V$ and
$\kappa_V$ to its SM expectation of unity~:
\begin{align}
\Delta\kappa_V=\kappa_V-1 \\
\Delta g^1_V=g^1_V-1
\end{align}
In almost all models the 
electromagnetic gauge invariance is taken for granted, such that
$\Delta g^1_\gamma$, the divergence of the W-charge from the unit
charge, is always zero. The parameter $\lambda_\gamma$ is
also set to zero in our analysis, since we are not aware 
of any computation of the dependence of $\epsilon_1$,  $\epsilon_2$ 
and  $\epsilon_3$ on $\lambda_\gamma$.

%
\section{Analysis and Results}

\begin{figure}[p]
\begin{center}
{\epsfig{file=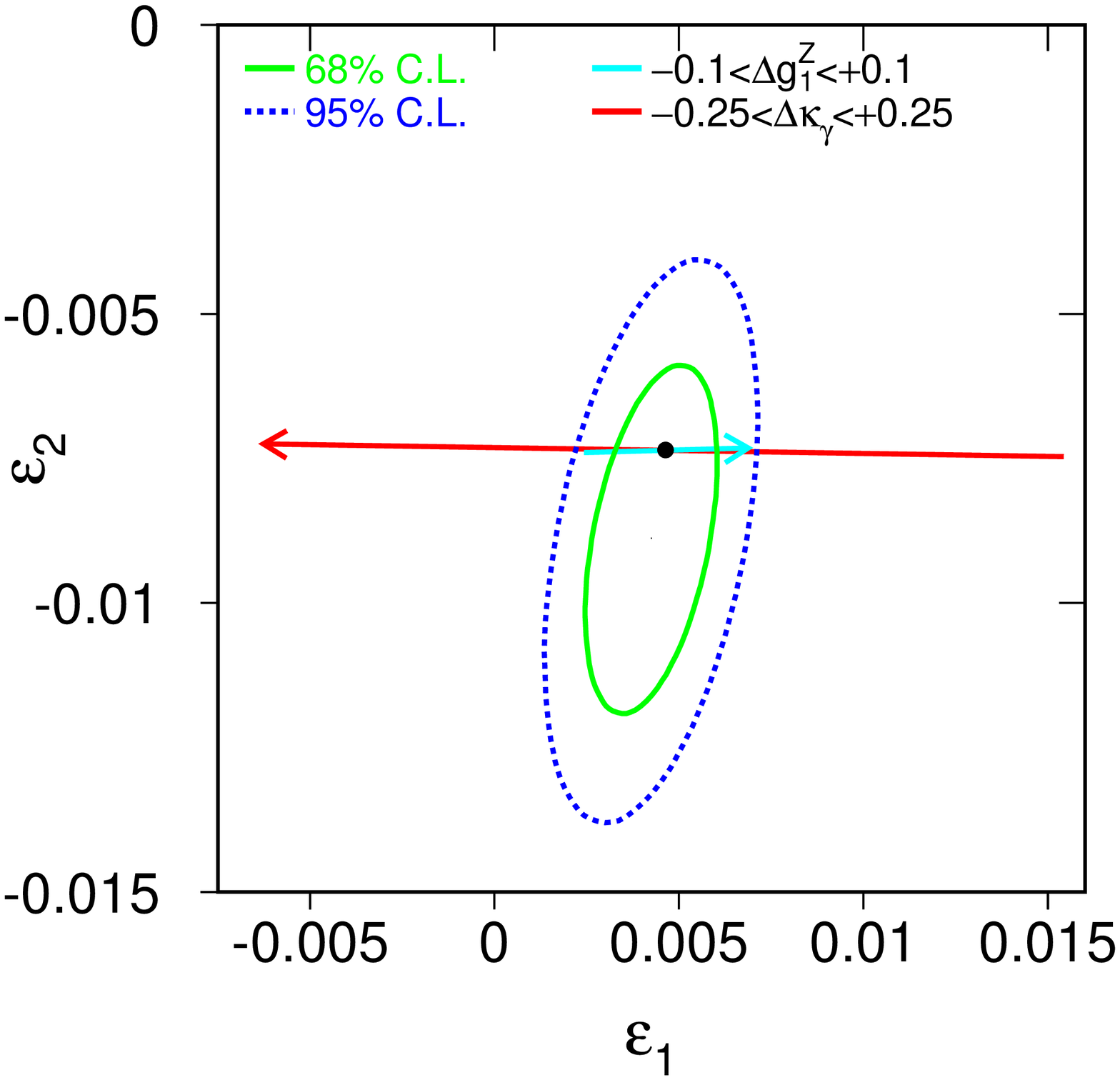,width=0.33\linewidth}}\hfill
{\epsfig{file=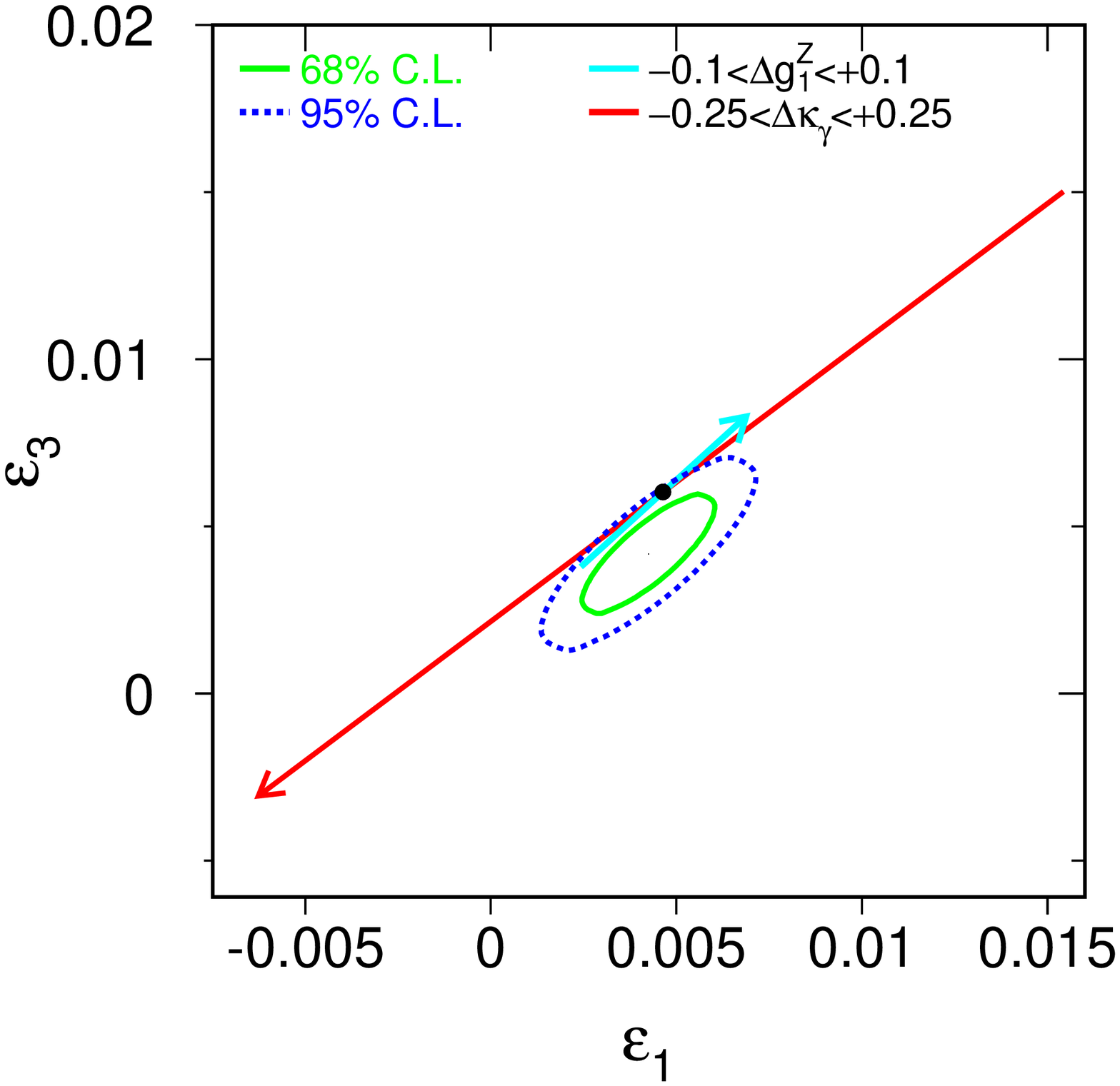,width=0.33\linewidth}}\\
{\epsfig{file=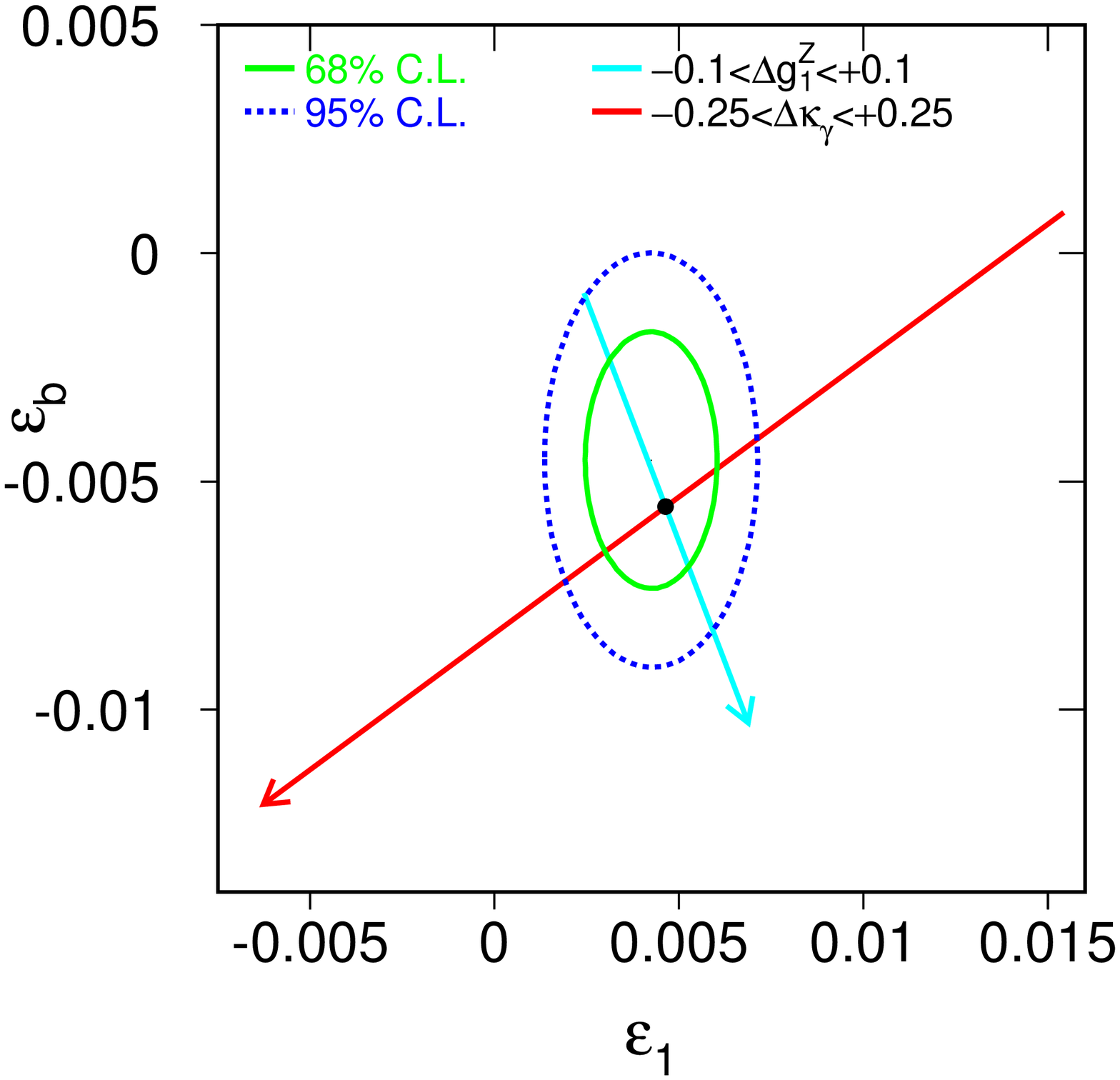,width=0.33\linewidth}}\hfill
{\epsfig{file=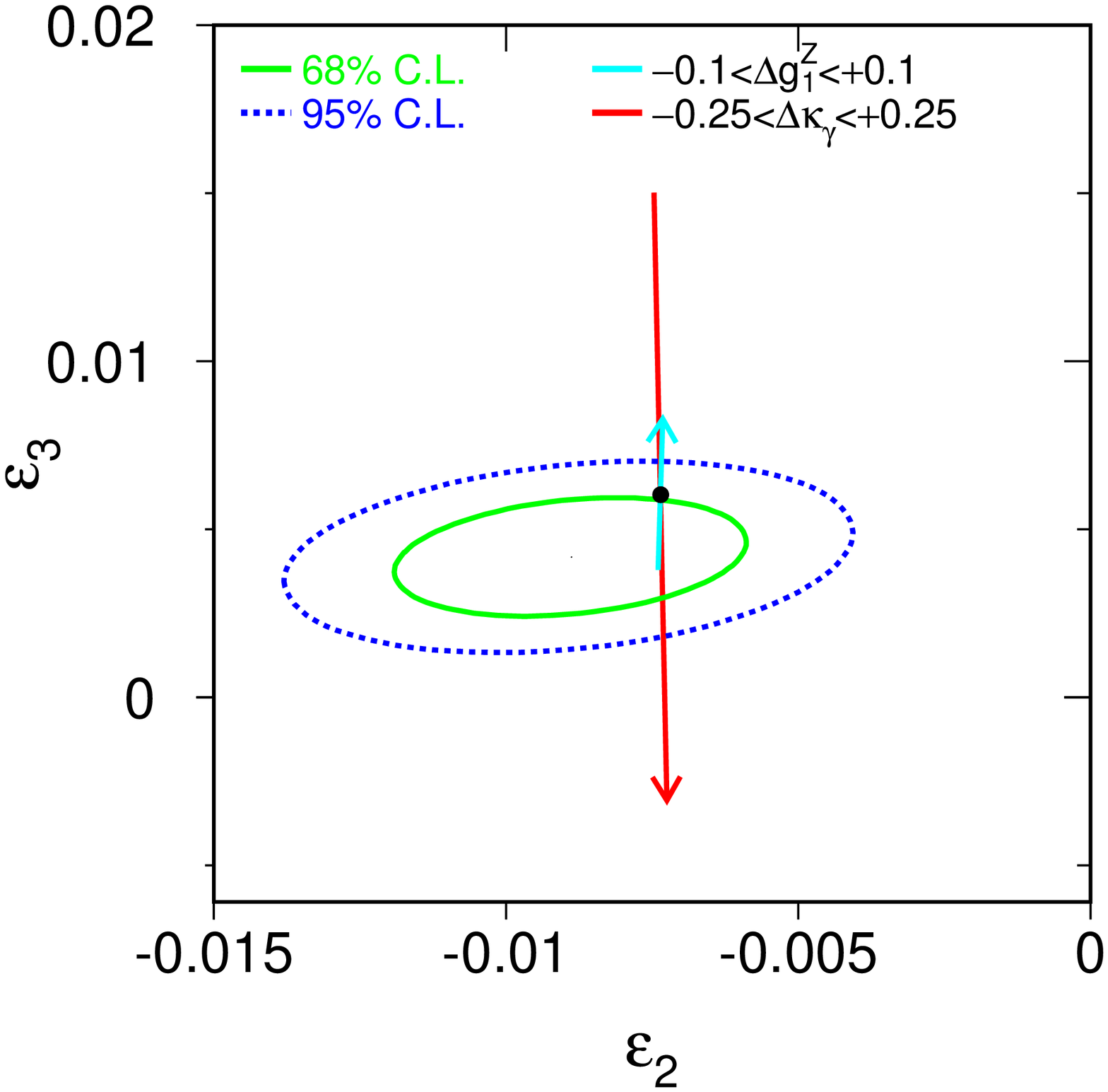,width=0.33\linewidth}}\\
{\epsfig{file=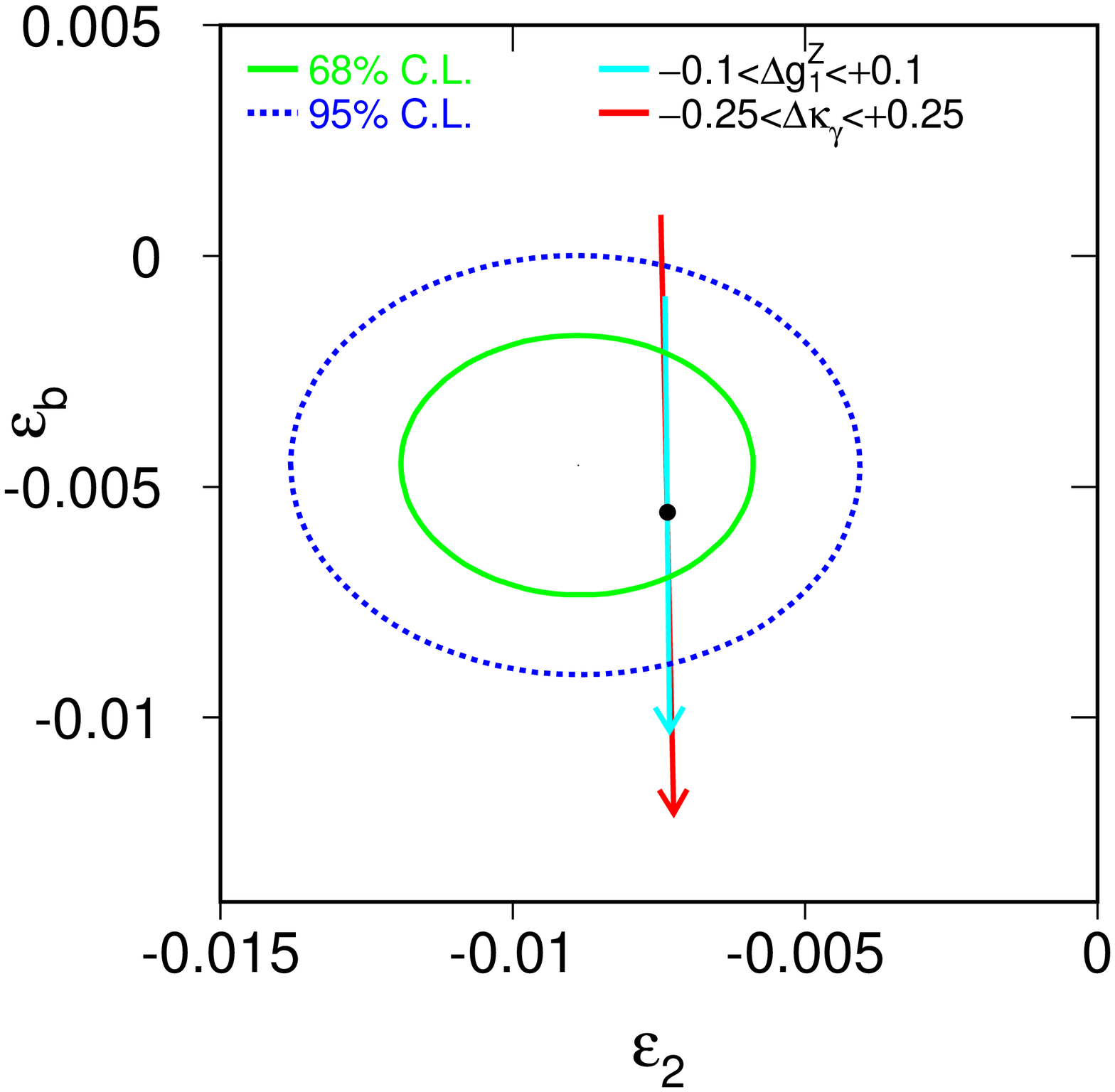,width=0.33\linewidth}}\hfill
{\epsfig{file=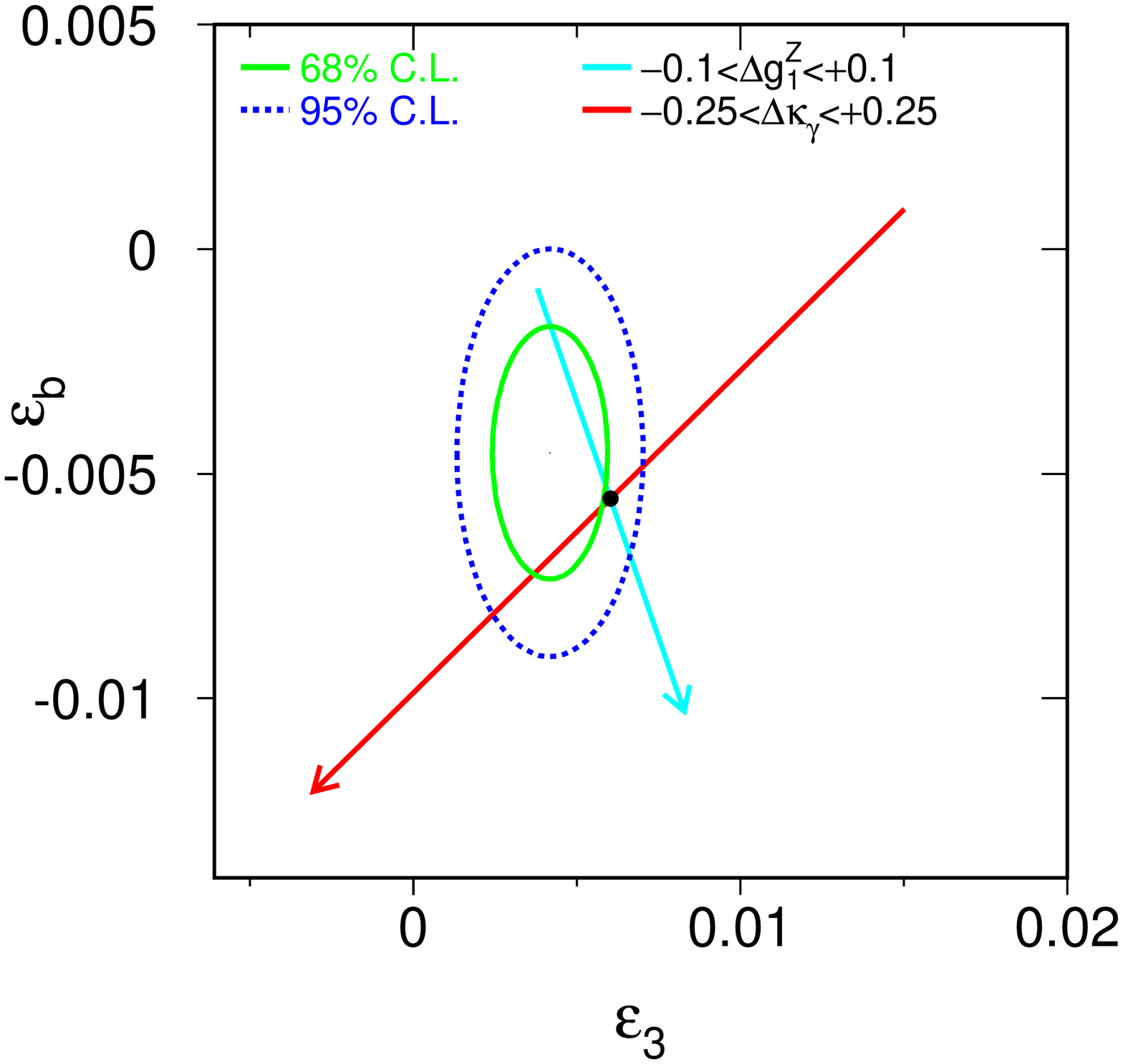,width=0.33\linewidth}}
\caption{The contours of the $\epsilon$ parameters. The arrows indicate
  the change of the SM prediction if the coupling parameters $\Delta
  g^1_Z$ and $\Delta\kappa_\gamma$ are varied according to the direct
  measurements of LEP 2 and TEVATRON.}
\label{fig:eps12}
\end{center}
\end{figure}

The preliminary measurements of electroweak parameters performed at LEP 1, SLAC
and TEVATRON are listed in table \ref{tab:ewpara}. The SM predictions
agree well with these measurements~\cite{karlgruen}. The analysis of this data set
proceeds via two steps. In the first step, the $\epsilon$ parameters
$\epsilon_1$, $\epsilon_2$, $\epsilon_3$ and
$\epsilon_b$~\cite{Altarelli:1991zd}:
\begin{align}
\epsilon_1&=\Delta\rho \\
\epsilon_2&=\cos^2\theta_W^0\Delta\rho+\frac{\sin^2\theta_W^0\Delta
  r_W}{\cos^2\theta_W^0-\sin^2\theta_W^0}-2\sin^2\theta_W^0\Delta k' \\
\epsilon_3&=\cos^2\theta_W^0\Delta\rho+(\cos^2\theta_W^0-\sin^2\theta_W^0)\Delta
k' \\
\epsilon_b&=\frac{g_A^b}{g_A^l}-1
\quad\hbox{and}\quad
\epsilon_b = 
\frac{g_V^b}{g_A^l}-\left(1-\frac{4}{3}(1+\Delta k')\sin^2\theta_W^0\right)
\end{align}
where:
\begin{eqnarray}
\sin^2\theta_W^0 & = & \frac{\pi\alpha(m^2_Z)}{\sqrt{2}G_F m^2_Z}
\end{eqnarray}
are extracted. These parameters are very sensitive to radiative
corrections and thus the influence of physics beyond the SM, hence
also very sensitive to non-SM TGCs. It is interesting to note that
$\epsilon_2$ and $\epsilon_b$ do not, on the one-loop level, depend on
the yet unknown Higgs-mass $m_H$.  Here $\Delta\rho$ stands for radiative
corrections to the $\rho$-parameter~\cite{Veltman:1977kh}, $\Delta r_w$ describes corrections to the
$G_F$-$M_W$ relation and $\Delta k'$ relates $\sin^2\theta_W^0$ to the effective
electroweak mixing angle~\cite{Altarelli:1991zd}.
As the fermion coupling constants depend on the $\epsilon$-parameters
one can extract these from the measurements reported in table
\ref{tab:ewpara} (except the top-quark mass), which all depend on
$g_V$, $g_A$ or $\sin^2\theta_W$.  A simultaneous fit to all four
parameters and in addition to the electromagnetic coupling constant
$\alpha_{em}(m_Z)$, the strong coupling constant $\alpha_s(m_Z)$ and
$m_Z$ gives the numbers quoted in table \ref{tab:epsilon}.  The
computation of the SM expectations shows that these values are in good
agreement with the measured ones, and they are also in good agreement
with other recent computations \cite{Altarelli:1998xf,hab_gruenewald}.
One finds strong correlations between $\epsilon_b$ and $\alpha_s$ as
well as for $\epsilon_1$ and $\epsilon_3$.  The latter is visible in
figure \ref{fig:eps12}, showing the two-dimensional contours of each
pair of $\epsilon$-parameters. These contour curves are compared with
the evolution of the $\epsilon$-parameters as a function of the TGC
coupling constants.

The dependence of the $\epsilon$-parameters on the WWV couplings is
shown in the following
equations\cite{Eboli:1994jh,Eboli:1998hb,hisz}~:

\begin{align}
-\frac{12\pi}{\alpha}\Delta\epsilon_1&=\left\{\left[\frac{27}{2}-\tan^2\theta_W\right]\frac{m_Z^2}{m_W^2}\ln\frac{\Lambda^2}{m_W^2}+\frac{9}{2}\frac{m_Z^2m_H^2}{m_W^4}\left[\ln\frac{\Lambda^2}{m_H^2}+\frac{1}{2}\right]\right\}\Delta\kappa_\gamma
\nonumber \\
&+\left\{\left[\tan^2\theta_W-\cot^2\theta_W\right]-\frac{9}{2}\frac{m_H^2}{m_W^2}\left[\ln\frac{\Lambda^2}{m_H^2}+\frac{1}{2}\right]\right\}\Delta
g^1_Z \label{equ:eps1} \\
\frac{12\pi}{\alpha}\Delta\epsilon_2&=\frac{m_Z^2}{m_W^2}\ln\frac{\Lambda^2}{m_W^2}\sin^2\theta_W\Delta\kappa_\gamma+ 
\cot^2\theta_W\ln\frac{\Lambda^2}{m_W^2}\Delta g^1_Z\\
\frac{12\pi}{\alpha}\Delta\epsilon_3&=\left\{\left[\cos^4\theta_W-7\cos^2\theta_W-\frac{3}{4}\right]\frac{m_Z^2}{m_W^2}\ln\frac{\Lambda^2}{m_W^2}-\frac{3}{4}\frac{m_H^2}{m_W^2}\left[\ln\frac{\Lambda^2}{m_W^2}+\frac{1}{2}\right]\right\}\Delta\kappa_\gamma\nonumber\\
&+\left\{10\cos^2\theta_W+\frac{3}{2}\right\}\ln\frac{\Lambda^2}{m_W^2}\Delta
  g^1_Z \\
\Delta\epsilon_b&=\frac{m_Z^2m_t^2}{64\pi^2m_W^4}\ln\frac{\Lambda^2}{m_W^2}\Delta\kappa_\gamma
\nonumber \\
&-\left[\frac{\cot^2\theta_W}{64\pi^2}\frac{m_Z^2m_t^2}{m_H^4}\ln\frac{\Lambda^2}{m_W^2}+\frac{3\cot^2\theta_W}{32\pi^2}\frac{m_t^2}{m_W^2}\ln\frac{\Lambda^2}{m_W^2}\right]\Delta
g^1_Z \label{equ:epsb}
\end{align}

These expressions are based on the constraints between TGCs quoted
earlier. All non-standard contributions are logarithmically divergent.
The coupling parameters, that are used here, are defined in dependence
on the new physics scale $\Lambda$ and a form factor f coming from the
new physics effect, eg.
\begin{align}
\Delta g^1_Z&=\frac{m_Z^2}{\Lambda^2}f.
\end{align}

Thus the coupling parameters vanish in the limit of a large new
physics scale, $\Lambda\rightarrow\infty$. The new physics scale in
the following measurement is set to 1~TeV. In addition a Higgs-mass of
300~GeV is assumed. 

\begin{figure}[p]
\vspace{-1cm}
\begin{center}
{\epsfig{file=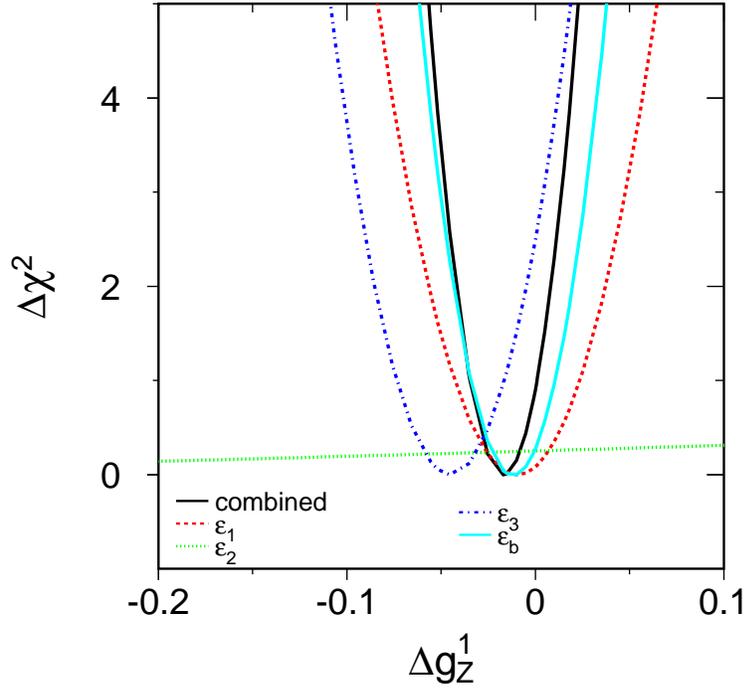,width=0.6\linewidth}}\\
{\epsfig{file=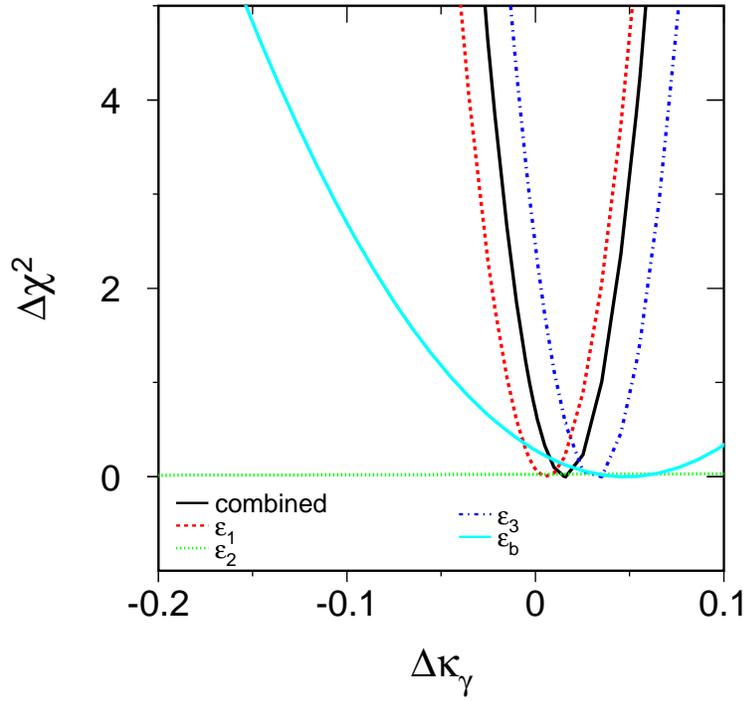,width=0.6\linewidth}}
\caption{The $\Delta\chi^2$ curves for the TGC couplings and the
  contributions of the different $\epsilon$ parameters. The combined
  curve is the add up of the single curves taking the correlation
  coefficients properly into account. The parameter $\epsilon_2$ has
  almost no sensitivity to TGCs.}
\label{fig:dg1z}
\end{center}
\end{figure}

A fit using equations \ref{equ:eps1} to \ref{equ:epsb} and the
difference of the measured values of the $\epsilon$-parameters and the
ones expected in the SM as shown in table~\ref{tab:epsilon} is used to
determine the TGC coupling parameters $\Delta g^1_Z$ and
$\Delta\kappa_\gamma$. The errors on the SM predictions of the
$\epsilon$-parameters are included, neglecting their correlations. 
The $\chi^2$ curves of a fit to each of these coupling constants,
setting the other to its SM value of zero, is shown in
figure~\ref{fig:dg1z}. One finds the following results:

\begin{align}
\Delta g^1_Z&=-0.017\pm0.018 \\
\mathrm{~~~~~~~~or}\nonumber  \\
\Delta\kappa_\gamma&=0.016\pm0.019   .
\end{align}

\begin{figure}[p]
\begin{center}
{\epsfig{file=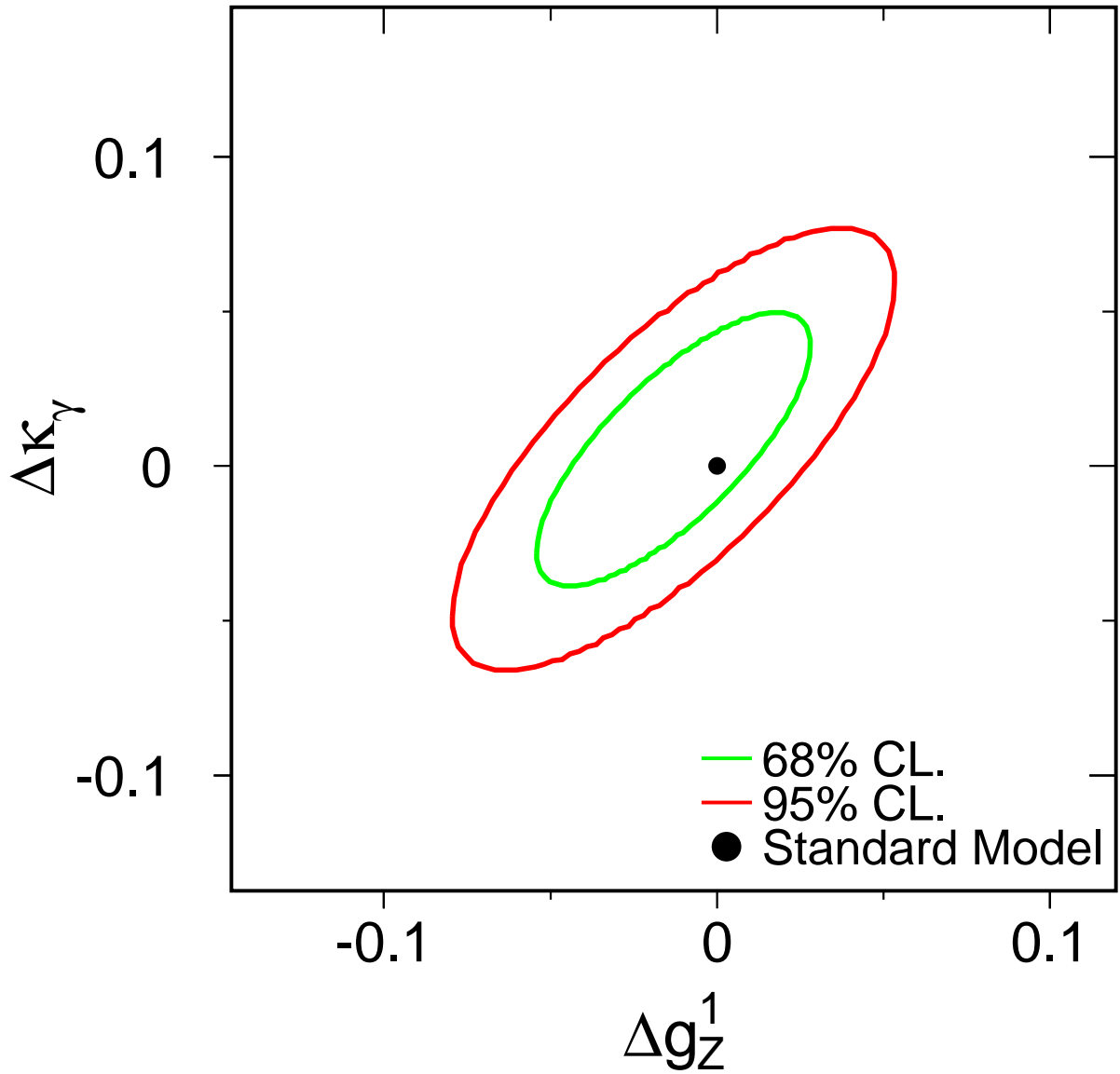,width=0.99\linewidth}}
\caption[]{The contour curves for the two dimensional fit, $\Delta
  g^1_Z$ versus $\Delta\kappa_\gamma$. The dot shows the SM expectation.}
\label{fig:dg1zdkg}
\end{center}
\end{figure}

If both couplings are allowed to vary in the fit, one finds the contour
plot in figure \ref{fig:dg1zdkg}. The corresponding numerical
values of the TGC-parameters are
\begin{align}
\Delta g^1_Z&=-0.013\pm0.027\nonumber \\
\Delta\kappa_\gamma&=0.005\pm0.029   ,
\end{align}
with a correlation of 75.5 percent.

The SM expectation of zero for both parameters agrees well with this
measurement. As 1~TeV is the lower limit of the new physics scale and
the couplings depend inversely on $\Lambda$, the errors decrease with
increasing $\Lambda$.  Higher Higgs masses decrease also the errors on
the TGC, while a lower Higgs mass increases the error.  Assuming a
100~GeV Higgs, the error on $\Delta g^1_Z$ increases by 1\% of the
error, while the
one-dimensional error on $\Delta\kappa_\gamma$ increases to 0.033.
The error of 5~GeV on $m_t$, as quoted in table \ref{tab:ewpara} 
has a negligible impact on the result.

The results presented above are more precise than recent direct
measurements of the LEP and TEVATRON collaborations \cite{karlgruen}:
$\Delta g^1_Z=0.00^{+0.12}_{-0.11}$ and
$\Delta\kappa_\gamma=0.28^{+0.33}_{-0.27}$. Here the parameters are
negatively correlated with -54 percent.  The direct measurement is
however more suitable for a general test of the TGCs while the
indirect measurement tests TGCs only in particular models.

Recent computations \cite{Eboli:1994jh,Eboli:1998hb} parametrise also
the dependence of $\epsilon_b$ on the coupling constants
$\lambda_\gamma$ and $g^5_Z$ giving access to a more general view of
the TGC couplings. Computations of the dependence of $\epsilon_1$,
$\epsilon_2$ and $\epsilon_3$ on the TGCs $\lambda_\gamma$ and $g^5_Z$
would be most useful to measure also these coupling constants more precisely.

%
%
\section{Acknowledgements}
We are very grateful to S. Riemann for bringing the possibility of the
indirect measurement of TGCs to our attention.  We thank F.Caravaglios
and G.Altarelli for clarifying discussions on the $\epsilon$
parameters 
and T. Hebbeker, W.Lohmann and T. Riemann for useful comments.

\clearpage

%
%
\bibliographystyle{../l3stylem}
\bibliography{../l3at183}

%
%
\newpage

\clearpage

\clearpage

\vspace{1cm}
\begin{table}[p]
\begin{center}
\begin{tabular}{||l|l|l||}
\hline
\hline
parameter      &  central value &    errors \\
 & & \\
\hline
\hline
 & & \\

$1/\alpha^{(5)}(m_Z)$ &  128.878       &    0.090 \\ 
$m_Z$          &   91.1867      & 0.0021 \\  
$\Gamma_Z$     &    2.4939      & 0.0024 \\  
$\sigma_\mathrm{had}$ &   41.491 & 0.058\\   
$R_e$          &   20.765       & 0.026  \\  
$A_{FB}^e$     &    0.01683     & 0.00096 \\ 
${\cal P}_e$   &    0.1479      & 0.0051  \\ 
${\cal P}_\tau$&    0.1431      & 0.0045 \\  
$\sin^2\theta_w^\mathrm{eff}(Q_\mathrm{fb})$& 0.2321  & 0.0010\\
$\sin^2\theta_w^\mathrm{eff}(A_\mathrm{LR})$& 0.23109 & 0.00029\\
$m_W$ (LEP2)          &   80.37        & 0.09\\
$m_W$ (p$\bar{\mathrm{p}}$)          &   80.41        & 0.09  \\
$R_b$          &    0.21656     & 0.00074 \\
$R_c$          &    0.1735      & 0.0044\\  
$A_{FB}^b$     &    0.0990      & 0.0021 \\
$A_{FB}^c$     &    0.0709      & 0.0044 \\
$A_b$     &    0.867       & 0.035  \\
$A_c$     &    0.647       & 0.040  \\
\hline
$m_t$          &    173.8       & 5.0 \\
 & & \\
\hline
\hline
\end{tabular}
\end{center} 
\caption[test]{Preliminary electroweak parameters that are used in the fit to the
  $\epsilon$ parameters. The correlations among the observables in the b
  and c quark sector as well as the one between $m_Z$, $\Gamma_Z$,
  $\sigma_\mathrm{had}$, $R_e$ and $A^e_{FB}$ is taken properly into
  account. Consult \cite{karlgruen} and references therein
  for details. $m_t$ is only used in the Standard Model calculation 
  of the $\epsilon$-parameters.}
\label{tab:ewpara}
\end{table}   

\clearpage

\begin{table}[p]
\begin{center}
\begin{tabular}{||l|r@{$\pm$}l|r|r|r|r|r|r|r|r||}
\hline
fit parameter            & \multicolumn{2}{|c|}{measured}   &   \mbox{~~~~MSM~~~}    &
\multicolumn{7}{|c||}{correlation matrix}\\
                         &        \multicolumn{2}{|c|}{}        &                    &   
$\frac{1}{\alpha^{(5)}}$   & $\alpha_s$ &  $m_Z$  &
$\epsilon_1$ &  $\epsilon_2$&
$\epsilon_3$ &  $\epsilon_b$  \\  
\hline
\hline
$1/\alpha^{(5)}(m_Z)$  &  128.878 & 0.090  &  \mbox{~~-~~~~~}  & 
{\bf 1.00} &  0.00  & 0.00 &   0.00 & -0.07 &  0.46 &  0.00 \\
$\alpha_s(m_Z)$              &  0.1244 & 0.0045 &     \mbox{~~~~-~~~~~}  &
0.00 &  {\bf 1.00}  & 0.00 & -0.45  & -0.22 & -0.31 &  -0.62  \\
$m_Z$                  &  91.1866 & 0.0021 &   \mbox{~~~~-~~~~~}  & 
0.00 &  0.00  & {\bf 1.00} & -0.06  & -0.01 & -0.02  &  0.00 \\
$\epsilon_1\times10^3$  &  4.2 & 1.2      &  $4.6\pm1.1$ &
0.00 & -0.45  & -0.06 &  {\bf 1.00} &  0.44 &  0.80 &  -0.01  \\
$\epsilon_2\times10^3$  &  $-8.9$ & 2.0     &   $-7.5\pm0.3$ &
-0.07 &  -0.22 & 0.00  & 0.44 & {\bf 1.00} &   0.26 &  -0.01 \\
$\epsilon_3\times10^3$  &  4.2 & 1.2      &   $5.8\pm0.7$ & 
0.46  &  -0.31 &  -0.02 &   0.80 &   0.26 &  {\bf 1.00} &  0.00 \\
$\epsilon_b\times10^3$  &  $-4.5$ & 1.9     & $-5.8\pm0.5$ &
0.00  &  -0.62 &  0.00 &  -0.01 &  -0.01 &  0.00 & {\bf 1.00} \\
\hline
\end{tabular}
\end{center}
\caption{The $\epsilon$ values in the SM and from a fit to the
 electroweak data summarised in table \ref{tab:ewpara}
 ($\chi^2/Ndf~=~11.6/11$, probability 39\%).}
\label{tab:epsilon}
\end{table}

\clearpage

\vspace{-1cm}
\clearpage

\clearpage

\end{document}